\documentclass{osa-article}
\usepackage[per-mode = symbol]{siunitx}
\usepackage{amsmath}
\usepackage{subfig}
\usepackage{multirow}
\usepackage{graphicx}
\usepackage{booktabs}
\usepackage{capt-of}
\usepackage{xcolor}
\usepackage[normalem]{ulem}

\journal{osac}

\articletype{Research Article}

\begin{document}
\title{Broadband low loss and ultra-low crosstalk waveguide crossings based on multimode interferometer for 840 nm operation}
\author{Stefan Nevlacsil,\authormark{1,*} Paul Muellner,\authormark{1} Martin Sagmeister,\authormark{2} Jochen Kraft,\authormark{2} and Rainer Hainberger\authormark{1}}
\address{\authormark{1}AIT Austrian Institute of Technology GmbH, Giefinggasse 4, 1210, Vienna, Austria\\
\authormark{2}ams AG, Tobelbader Stra\ss e 30, 8141 Premst\"atten, Austria
}
\email{\authormark{*}stefan.nevlacsil.fl@ait.ac.at} 
%
\begin{abstract}
Broadband low loss and ultra-low crosstalk waveguide crossings are a crucial component for photonic integrated circuits to allow a higher integration density of functional components and an increased flexibility in the layout. We report the design of optimized silicon nitride waveguide crossings based on multimode interferometer structures for intersecting light paths of TE/TE-like, TM/TM-like and TE/TM-like polarized light in the near infrared wavelength region of \SIrange{790}{890}{\nano\metre}. 
The crossing design for diverse polarization modes facilitates dual polarization operation on a single chip. For all configurations the loss of a single crossing was measured to be \textasciitilde \SI{0.05}{dB} at \SI{840}{\nano\metre}. Within the \SI{100}{\nano \metre} bandwidth losses stayed below \SI{0.16}{dB}. The crosstalk was estimated to be on the order of \SI{-60}{dB} by means of 3D finite difference time domain simulations.
\end{abstract}
%
\section{Introduction}
Waveguide crossings are a key building block for the realization of photonic integrated circuits (PICs) with higher functional complexity e.g.\ in photonic networks \cite{Shacham.2007,Sherwood-Droz.2013,Dumais.2018}. This holds especially true for PICs that require end facet coupling, which inherently sets limitations in the layout options. In principle, one could gain significant design freedom for the PIC layout by using waveguide grating couplers, which can be placed arbitrarily on the chip. However, this positioning flexibility is obtained at the expense of higher coupling losses, stronger wavelength dependence, and inherent polarization dependence. The coupling efficiency of gratings can only be improved by significant additional efforts in fabrication. Moreover, the strong wavelength dependence limits the applicability for broadband applications such as optical coherence tomography (OCT). 
On the other hand, end facet coupling via inverted tapers can achieve coupling efficiency of close to \SI{-1}{dB} over a bandwidth of \SI{100}{\nano \metre} \cite{Hainberger.02.02.2019}. In addition, inverted tapers with a squared-shaped cross section are polarization invariant. 
The use of waveguide crossings mitigates the limitations in on-chip routing and allows taking full advantage of the benefits of end facet coupling. However, to avoid impairments in signal performance the waveguide crossings have to feature low losses and ultra-low crosstalk. Furthermore, depending on the application, broadband operation has to be ensured.
\begin{table}[htbp]
\centering
\resizebox{0.92\textwidth}{!}{%
\begin{tabular}{|c|c|c|c|c|c|c|c|}
\hline
Type & Detail & Year & Material & $\lambda_C$ [nm] & BW [nm] & IL [dB] & CT [dB] \\ \hline
\multirow{3}{*}{\shortstack{Single\\mode\\waveguide}} & Angled & \cite{Sanchis.2007} (2007) & SOI & 1550 & 35 & $~1.6^{\,\text{M}}$ & $<-20^{\,\text{M}}$ \\ \cline{2-8} 
 & Offset & \cite{Tanaka.2009} (2009) & SOI & 1550 & - & $0.02^{\,\text{S,*}}$ & $-55^{\,\text{S,*}}$ \\ \cline{2-8} 
 & Grating & \cite{Bock.28.06.200902.07.2009} (2010) & SOI & 1550 & 60 & $<0.04^{\,\text{M,*}}$ & $<-40^{\,\text{M,*}}$ \\ \hline
\multirow{5}{*}{\shortstack{Mode\\expander}} & Elliptical & \cite{Fukazawa.2004} (2004) & SOI & 1550 & 60 & $<0.10^{\,\text{M}}$ & $<-25^{\,\text{M}}$ \\ \cline{2-8} 
 & Parabolic& \cite{Bogaerts.2007} (2007) & SOI & 1550 & 40 & $<0.16^{\,\text{M}}$ & $<-40^{\,\text{M}}$ \\ \cline{2-8} 
 & Algorithm & \cite{Sanchis.2009} (2009) & SOI & 1550 & 20 & $<0.20^{\,\text{M}}$ & $<-40^{\,\text{S}}$ \\ \cline{2-8} 
 & Elliptical & \cite{Shinobu.2010} (2010) & SOI & 1550 & - & $~0.24^{\,\text{M}}$ & - \\ \cline{2-8} 
 & Parabolic& \cite{Xu.2011} (2011) & SOI & 1550 & 40 & $~0.44^{\,\text{S}}$ & $<-40^{\,\text{S}}$ \\ \hline
\multirow{19}{*}{MMI} & \multirow{9}{*}{\shortstack{Single\\taper}} & \cite{Liu.2004} (2004) & $\Delta$n=2\% & 1550 & 40 & $<0.05^{\,\text{S}}$ & - \\ \cline{3-8} 
 &  & \cite{Chen.2006} (2006) & SOI & 1550 & 60 & $<0.50^{\,\text{M}}$ & $<-30^{\,\text{S}}$ \\ \cline{3-8} 
 &  & \cite{Xu.2008} (2008) & SOI & 1550 & 100 & $<0.12^{\,\text{S}}$ & $<-40^{\,\text{S}}$ \\ \cline{3-8} 
 &  & \cite{Chiu.2010} (2010) & SOI & 1550 & 100 & $<0.40^{\,\text{S}}$ & $<-37^{\,\text{S}}$ \\ \cline{3-8} 
 &  & \cite{Xu.2013} (2013) & SOI & 1550 & 24 & $<0.09^{\,\text{S,a}}$ & $<-32^{\,\text{S,a}}$ \\ \cline{3-8} 
 &  & \cite{Li.2014} (2014) & SOI & 1550 & 60 & $<0.11^{\,\text{M}}$ & $<-40^{\,\text{M}}$ \\ \cline{3-8} 
 &  & \cite{Xu.2016} (2016) & SOI & 1560 & 80 & $<1.5^{\,\text{M,b}}$ & $<-18^{\,\text{M,b}}$ \\ \cline{3-8} 
 &  & \cite{Celo.2017} (2017) & SOI & 1550 & 40 & $<0.04^{\,\text{M}}$ & $<-40^{\,\text{M}}$ \\ \cline{3-8} 
 &  & \cite{Yang.2019} (2019) & SiN & 1580 & 120 & $<0.70^{\,\text{M,*}}$ & $<-42^{\,\text{M,*}}$ \\ \cline{2-8} 
 & \multirow{4}{*}{\shortstack{Multi\\section\\taper}} & \cite{Chen.19.04.201221.04.2012} (2012) & SOI & 1550 & 100 & $<0.37^{\,\text{S}}$ & $<-35^{\,\text{S}}$ \\ \cline{3-8} 
 &  & \multirow{2}{*}{\cite{Ma.2013} (2013)} & \multirow{2}{*}{SOI} & 1310 & 60 & $<0.05^{\,\text{M}}$ & $<-37^{\,\text{M}}$ \\ \cline{5-8} 
 &  &  &  & 1550 & 60 & $<0.07^{\,\text{M}}$ & $<-37^{\,\text{M}}$ \\ \cline{3-8}
 &  & \cite{Dumais.2017} (2017) & SOI & 1550 & 35 & $<0.03^{\,\text{M}}$ & $<-40^{\,\text{M}}$ \\ \cline{2-8} 
 & \multirow{3}{*}{Algorithm} & \cite{Chang.2018} (2018) & SOI & 1560 & 60 & $<0.60^{\,\text{M,b}}$ & $<-24^{\,\text{M,b}}$ \\ \cline{3-8} 
 &  & \cite{Yu.2019} (2019) & SOI & 1540 & 200 & $<0.20^{\,\text{M,*}}$ & $<-28^{\,\text{M,*}}$ \\ \cline{3-8} 
 &  & \cite{Hoffman.2019} (2019) & SOI & 1550 & 100 & $<0.15^{\,\text{M}}$ & $<-30^{\,\text{M}}$ \\ \cline{2-8} 
 & Angled & \cite{Kim.2014b} (2014) & SOI & 1560 & 60 & $<0.30^{\,\text{M,*}}$ & $<-38^{\,\text{M,*}}$ \\ \cline{2-8} 
 & \multirow{2}{*}{Array} & \cite {Zhang.2013} (2013) & SOI & 1565 & 90 & $<0.02^{\,\text{M}}$ & $<-40^{\,\text{M}}$ \\ \cline{3-8} 
 &  & \cite{Liu.2014} (2014) & SOI & 1550 & 100 & $<0.10^{\,\text{M}}$ & $<-25^{\,\text{M}}$ \\ \hline
\multicolumn{2}{|c|}{\multirow{7}{*}{Non-planar}} & \cite{Tsarev.2011} (2011) & Si/SU-8 & 1550 & 100 & $<0.36^{\,\text{S}}$ & $<-60^{\,\text{S}}$ \\ \cline{3-8} 
\multicolumn{2}{|c|}{} & \cite{Wakayama.2011} (2011) & Si/SiOx & 1550 & - & $<0.68^{\,\text{S}}$ & $-36^{\,\text{S}}$ \\ \cline{3-8} 
\multicolumn{2}{|c|}{} & \cite{Jones.2013} (2013) & Si/SiN & 1550 & 100 & $<0.20^{\,\text{M,*}}$ & $<-49^{\,\text{M,*}}$ \\ \cline{3-8} 
\multicolumn{2}{|c|}{} & \cite{Moreira.2013} (2013) & SiN & 1560 & 80 & $<1.10^{\,\text{M}}$ & - \\ \cline{3-8} 
\multicolumn{2}{|c|}{} & \cite{Shang.2015} (2015) & SiN & 1550 & - & $0.17^{\,\text{M}}$ & $<-52^{\,\text{M}}$ \\ \cline{3-8} 
\multicolumn{2}{|c|}{} & \cite{Chiles.2017} (2017) & SOI & 1535 & 70 & $<0.4^{\,\text{M}}$ & $<-35^{\,\text{M}}$ \\ \cline{3-8} 
\multicolumn{2}{|c|}{} & \cite{Sacher.2017} (2017) & Si/SiN & 1550 & 140 & $<0.3^{\,\text{M}}$ & $<-56^{\,\text{M}}$ \\ \hline
\end{tabular}%
}
\caption{Different types of Si and SiN based waveguide crossings reported in literature. Superscripts "M" and "S" denote measured and simulated results, respectively. Further denoted are crossings for slot waveguides (a), higher modes (b), and crossings for which values of both polarizations are available with TE-like shown (*). Abbreviations: BW~=~bandwidth, IL~=~insertion loss, CT~=~crosstalk.}
\label{tab:prior_art}
\end{table}
In literature, different types of waveguide crossings have been reported.  Table~\ref{tab:prior_art} provides an overview on these types. Design approaches can be divided into four major categories using single mode waveguides, mode expander, multimode interferometers (MMI), and non-planar multi-layer waveguides.
\newpage
Some of the earlier attempts to achieve low loss waveguide crossings use single mode waveguides with non orthogonal intersections \cite{Sanchis.2007}, including an offset \cite{Tanaka.2009} or with subwavelength grating structures \cite{Bock.28.06.200902.07.2009}. The angled waveguide approach exhibited comparatively high losses and the offset type was designed only for a single wavelength. The subwavelength grating based waveguide crossing showed a promising performance, however, these structures have high fabrication demands.

Other designs rely on the expansion of modes in a mode expander to reduce the diffraction of the mode at the transition to the slab waveguide of the intersection. The expansion of the mode is done without excitation of higher order modes by an appropriate choice of the taper form e.g.\ parabolic \cite{Fukazawa.2004}. The downside of these crossings is that some require additional shallow etching steps \cite{Bogaerts.2007,Xu.2011} or have comparatively small bandwidths.

MMI crossings are the most frequently used type because they can be easily fabricated and have the potential for high efficiency and broadband operation. In MMI crossings, higher order modes are excited in a multimode waveguide to achieve low loss intersections. To avoid excessive losses from the transition of the narrow single mode to the broader multimode waveguide section tapers are typically employed, which can be divided into standard single function tapers (single taper), e.g.\ linear and Gaussian tapers, and tapers with multiple expansion sections (multi section taper). More uncommon single taper MMI crossing designs are used for slot waveguides \cite{Xu.2013} and higher order modes \cite{Xu.2016}. Recent examples of improved MMI crossings employ special algorithms to obtain subwavelength structures \cite{Chang.2018} or optimize the general shape \cite{Yu.2019,Hoffman.2019}. Other approaches to reduce losses and crosstalk use non orthogonal intersecting MMI waveguides \cite{Kim.2014b} or an array of intersections in a single MMI waveguide to create low loss Bloch waves \cite{Zhang.2013,Liu.2014}.

Non-planar waveguide crossings represent another type of waveguide crossing that rely on multiple waveguide layers to avoid directly intersecting waveguides. There have been approaches where light is transferred between waveguides made of either the same material \cite{Moreira.2013,Shang.2015,Chiles.2017} or of two different materials with silicon being one of them and a polymer \cite{Tsarev.2011}, SiO$_\text{x}$ \cite{Wakayama.2011} or silicon nitride \cite{Jones.2013,Sacher.2017} being the other. These designs exhibit very low direct crossing losses and crosstalk e.g.\ \SI{0.002}{dB} and <\SI{-56}{dB} \cite{Sacher.2017}, respectively. However, the inter-layer transition loss is a major contribution to the overall insertion loss (IL) of the crossings given in table~\ref{tab:prior_art}. Furthermore, these designs are significantly more difficult to fabricate with multiple layers having to be fabricated at very precise distances.

The vast majority of the waveguide crossings described in literature have been realized on a silicon on insulator (SOI) waveguide platform and operate at the telecom wavelengths \SI{1550}{\nano \metre} or \SI{1310}{\nano \metre}. Recently, Yang et al. \cite{Yang.2019} reported an MMI crossing of double strip silicon nitride (SiN) waveguides surrounded by silicon dioxide (SiO$_\text{2}$) cladding material operating at \SI{1550}{\nano \metre}. In the case of non-planar crossings two examples can be found where only SiN is used as waveguiding material \cite{Moreira.2013,Shang.2015}. For biological applications wavelengths in the visible to near infrared region below \SI{1}{\micro \metre} are of strong interest due to the much high water absorption at the telecom wavelengths. For such applications SOI can no longer be used because of the strong light absorption in silicon at these wavelengths \cite{Subramanian.2013}. For a wavelength of \SI{850}{\nano\metre} polymer based waveguide crossings have been reported for the use in optical interconnects \cite{Bamiedakis.07,Swatowski.14,Ishigure.15,Kinoshita.16,Abe.18}. The low refractive index contrast in these polymer waveguides allows for very low losses below \SI{0.01}{dB/crossing}. However, the low index contrast comes with the trade-off that bends have to be comparatively large to avoid excessive radiative losses, which causes large PIC dimensions. A good compromise can be found in SiN as waveguide material. SiN waveguides with SiO$_\text{2}$ cladding have a moderate refractive index contrast with more manageable minimum bend radii. A variety of SiN integrated optical waveguide key components operating around \SI{840} {\nano \metre} wavelength have already been demonstrated \cite{Subramanian.2013, Martens.2018}. Furthermore, SiN has the advantage that it can be fabricated in a CMOS compatible way using plasma enhanced chemical vapour deposition (PECVD) with equipment readily available in foundries  \cite{Gorin.2008,Subramanian.2013,Muellner.2015,Martens.2018,Porcel.2019}. This allows a monolithic integration with optoelectronics and electronics resulting in more compact and robust integrated circuits.

In our design we chose MMI type waveguide crossings because of their high performance and low fabrication requirements. We use PECVD SiN for the waveguide crossings to allow CMOS compatible fabrication and a wavelength operation window of \SIrange{790}{890}{\nano\metre}, which is to our knowledge shown for the first time. To facilitate different PIC designs in biological applications the MMI crossings in this work are designed for single polarization operation, i.e.\ only transverse electric (TE)-like or transverse magnetic (TM)-like polarization and in dual polarization operation by enabling intersections of waveguide modes with TE-like and TM-like polarization. A crossing for light with TE/TM-like polarization has been shown before with non orthogonal intersecting MMI sections \cite{Kim.2014b}. The taper width and the MMI width were specified for each polarization, and the angle and MMI lengths were optimized. In our design, the MMI sections are kept perpendicular to each other to avoid higher crosstalk in one of the cross ports. The taper and MMI length are optimized for each polarization to achieve optimal performance. These mixed polarization crossings can be employed to build for example a 2x2 polarization-diversity switch \cite{Kim.2014}. The presented switch design by Kim et al. uses a much easier to implement polarization beam splitter compared to other designs which rely on a polarization rotator. Since we have reported the design of a polarization beam splitter in a previous work \cite{Hainberger.02.02.2019}, such a switch can readily be implemented for the wavelength range of \SIrange{790}{890}{\nano\metre} with the waveguide crossing described in this work.
\section{Multimode interferometer crossing design}
\begin{figure}
\centering
\includegraphics[width=13.3cm]{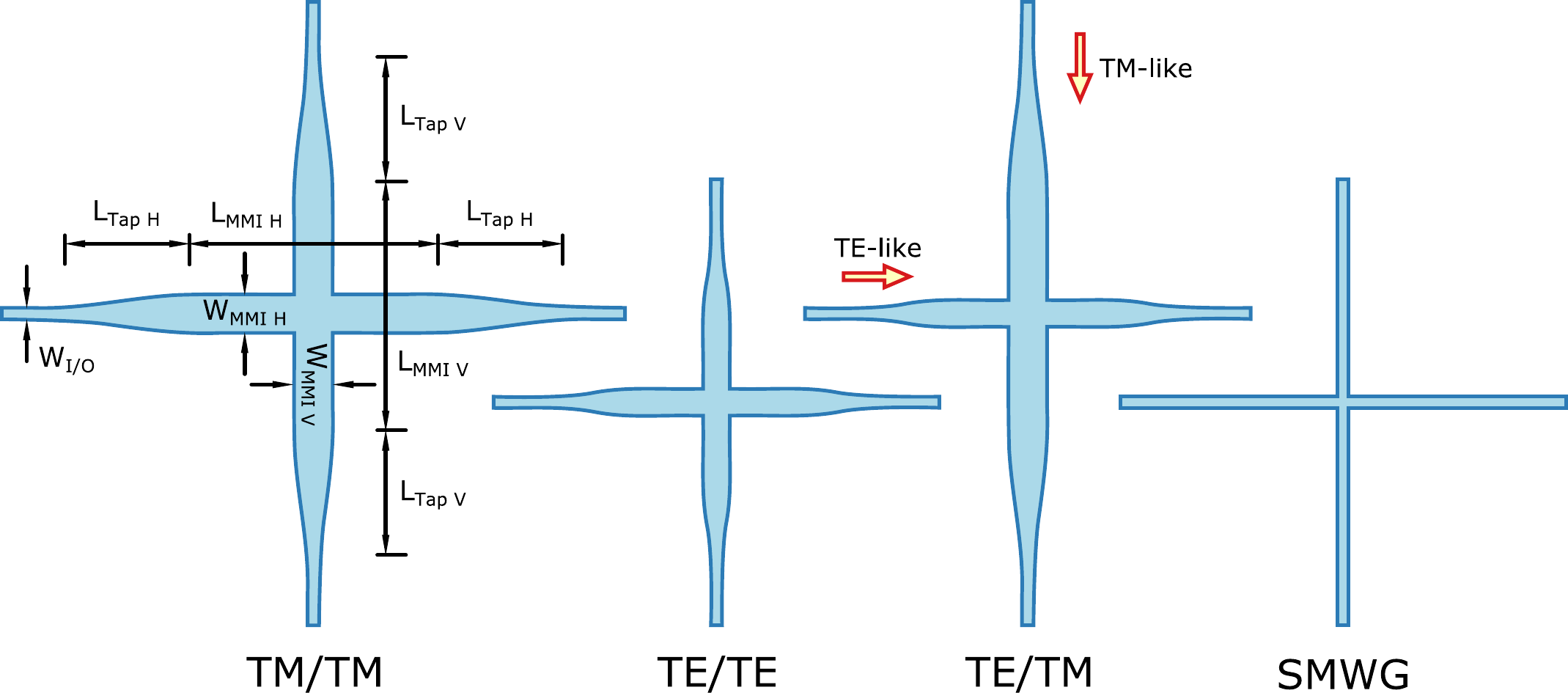}
\caption{Schematics of tested waveguide crossings with corresponding geometry parameters exemplarily shown for the TM/TM type. The TE/TM type allows the low-loss crossing of light with different polarizations. The input waveguides have a width of \SI{0.7}{\micro\metre} for all crossing types. The last schematic depicts a crossing consisting of single mode waveguides (SMWG) without an MMI.}
\label{fig:crossing_types}
\centering
\captionof{table}{Values of the geometry parameters for the different crossing types.}
\begin{tabular}{cccccccc} \toprule
Type & W$_\text{I/O}$  & W$_\text{MMI H}$ & W$_\text{MMI V}$ & L$_\text{MMI H}$ & L$_\text{Tap H}$ & L$_\text{MMI V}$ & L$_\text{Tap V}$ \\ \midrule
TM/TM & 0.7 & 2.3    & 2.3     & 18.04   & 9.87   & 18.04  & 9.87   \\
TE/TE & 0.7 & 1.65    & 1.65     & 9.15   & 3.82   & 9.15  & 3.82   \\
TE/TM & 0.7 & 1.65    & 2.3     & 9.15   & 3.82   & 18.04  & 9.87   \\
SMWG & 0.7 & -    & -    & 0   & 0   & 0  & 0   \\
\bottomrule \\
\end{tabular}
\label{tab:crossing_parameters}
\end{figure}

Waveguide crossings consisting of MMIs rely on the self-imaging principle \cite{Soldano.1995}. Self-imaging is the periodic recurrence of the input field distribution at certain positions in the MMI due to a superposition of the available modes. This input mode field is significantly more confined compared to the individual modes of the MMI. Therefore, at the position where the field distribution is more compact, any disturbance at the sidewalls, i.e.\ a crossing waveguide, in the ideal case does not affect the propagating light. The period of the recurrence is given by the so called beating length $\text{L}_\text{B}$. In the first step of the MMI design, the optimal width has to be determined for both the TE-like and the TM-like polarization. With increasing width the length of the multimode interferometer also increases \cite{Soldano.1995}, which is detrimental for both the footprint of the device and the distance the light has to propagate in the intersecting slab waveguide before reentering the MMI section. Therefore, in our designs the widths of the MMIs were kept as narrow as possible for both polarizations. In the crossing design, shown in Fig.~\ref{fig:crossing_types}, the single mode waveguide excites the modes in the MMI at the center of the MMI input. Due to this symmetric configuration the mode field of the fundamental even mode of the single mode input waveguide (TE$_0$/TM$_0$) can only excite the even modes of the MMI (TE$_{0,2,4,\dots}$/TM$_{0,2,4,\dots}$). Therefore, the width of the MMI section was chosen to enable the first higher even mode (TE$_2$/TM$_2$) resulting in a beating of the two modes in the MMI (TE$_{0,2}$/TM$_{0,2}$).
To find the optimum width, eigenmode simulations were carried out with the tool MODE \cite{LumericalInc..2018}, which employs the finite difference method. For the crossings, SiN wire waveguides are used with a height of \SI{160}{\nano \metre}. This height is well suited for the realization of different integrated optical waveguide functional key components \cite{Hainberger.02.02.2019, Hainberger.01.04.2019, Nevlacsil.2018}. A single mode waveguide width of \SI{700}{\nano \metre} was chosen to ensure single mode operation, while keeping the confinement to a maximum to facilitate small bend radii. The refractive indices for both polarizations of the material were determined with ellipsometry at a wavelength of \SI{840}{\nano \metre} and amounted to 1.916 for SiN and 1.455 for SiO$_\text{2}$, respectively. With the results from the eigenmode simulations the widths were chosen as \SI{1.65}{\micro \metre} for TE-like polarization and \SI{2.3}{\micro \metre} for TM-like polarization both with a safety margin of \textasciitilde \SI{0.2}{\micro \metre} before the higher order symmetric mode is no longer supported for \SI{890}{\nano \metre}. The next step was the design of the MMI crossing. Gaussian tapers were used from the single mode waveguides to the MMI section. This taper type was reported to have the smallest variation across the wavelength bandwidth for the insertion loss compared to other taper types as radical, linear and quadratic \cite{Chiu.2010}. Furthermore, this taper type has a comparatively small footprint. To relax fabrication requirements the taper end width was set equal to the MMI width resulting in a step free transition. In order to increase the reproducibility in fabrication the sharp edges of the intersecting perpendicular waveguides were rounded to a minimum fabrication radius of \SI{200}{\nano \metre}.

For the optimization of the crossing length a value for the beat length can be calculated by using the propagation constants from the MMI eigenmode simulations with $\text{L}_\text{B} = 2\pi/(\beta_0 - \beta_2)$ \cite{Chiu.2010}. With the first reoccuring self image after $\text{L}_\text{B}$ at the position of the intersection the total MMI length is then approximately $2\text{L}_\text{B}$. For TE-like polarization the value of $2\text{L}_\text{B} \approx \SI{14}{\micro\metre}$ and for TM-like polarization $2\text{L}_\text{B} \approx \SI{29}{\micro\metre}$. Due to the non-adiabatic nature of the Gaussian taper higher order modes are also excited in the taper and the beat length is partially provided by the taper. The final optimization was done with 3D finite difference time domain (FDTD) simulations \cite{LumericalInc..2018}. In contrast to other methods such as the eigenmode expansion or the beam propagation method, the FDTD method is well suited for the simulation of crossings due to its capability to accurately describe abrupt changes in the cross section of the waveguide. In the FDTD simulation tool, the geometry of the crossing structure, as seen in Fig.~\ref{fig:crossing_types}, was fully parameterized with the internal programming language, which allowed the usage of Lumerical's particle swarm optimization algorithm. The mean transmission over the bandwidth of \SIrange{790}{890}{\nano\metre} was chosen as figure of merit. This value was maximized by changing both the length of the taper and the length of the MMI section in the TE/TM-like polarization configuration. Table \ref{tab:crossing_parameters} summarizes the resulting values of the geometry parameters defining the MMI crossing geometries. 
\begin{figure}
    \centering
    \includegraphics[width=13.3cm]{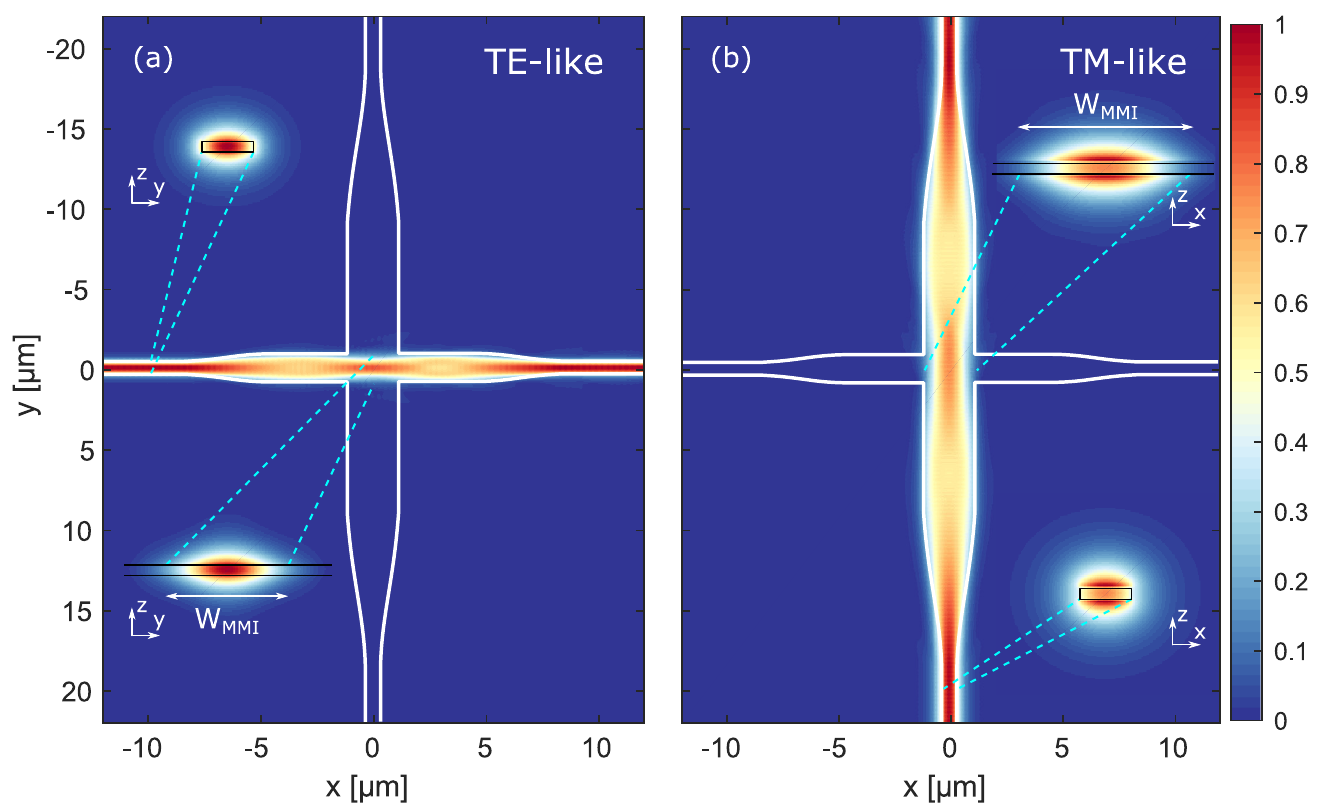}
    \caption{Normalized field amplitude of the 3D-FDTD simulations in the TE/TM crossing for (a)~TE-like polarized light propagating from left to right (b)~TM-like polarized light propagating from top to bottom. The insets depict the normalized field amplitude distribution at the indicated positions in the single mode waveguide compared to the MMI. At the crossing position the mode expansion is reduced due to the beating behaviour of the MMI.}
    \label{fig:MMI_Crossing}
\end{figure}
Figure~\ref{fig:MMI_Crossing} shows the normalized field amplitudes of the final design for the TE/TM crossing calculated by the 3D-FDTD simulation.

\section{Measurement and simulations}
\begin{figure}
    \centering
    \includegraphics[width=13.3cm]{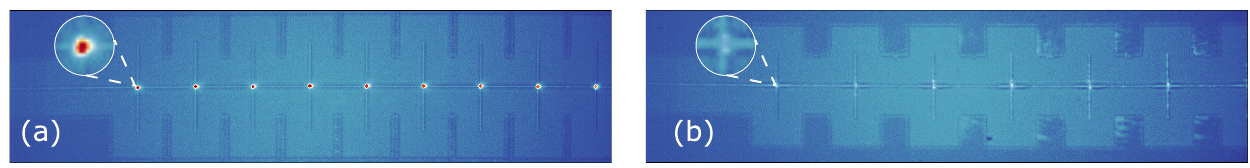}
    \caption{False color image of cascaded crossings recorded with an infrared camera with light of \SI{840}{\nano \metre} wavelength propagating in TM-like polarization for (a)~a standard SMWG crossing and (b)~the optimized TE/TM crossing. The images were recorded with identical camera settings. The distinct emission spots at the SMWG crossings indicate the higher losses in comparison to the MMI crossings. The stripes between the waveguide structures originate from the light scattered at lithographic fill patterns that are used to facilitate a homogeneous fabrication.}
    \label{fig:microscopy_image}
\end{figure}

\begin{figure}
    \centering
    \includegraphics[width=13.3cm]{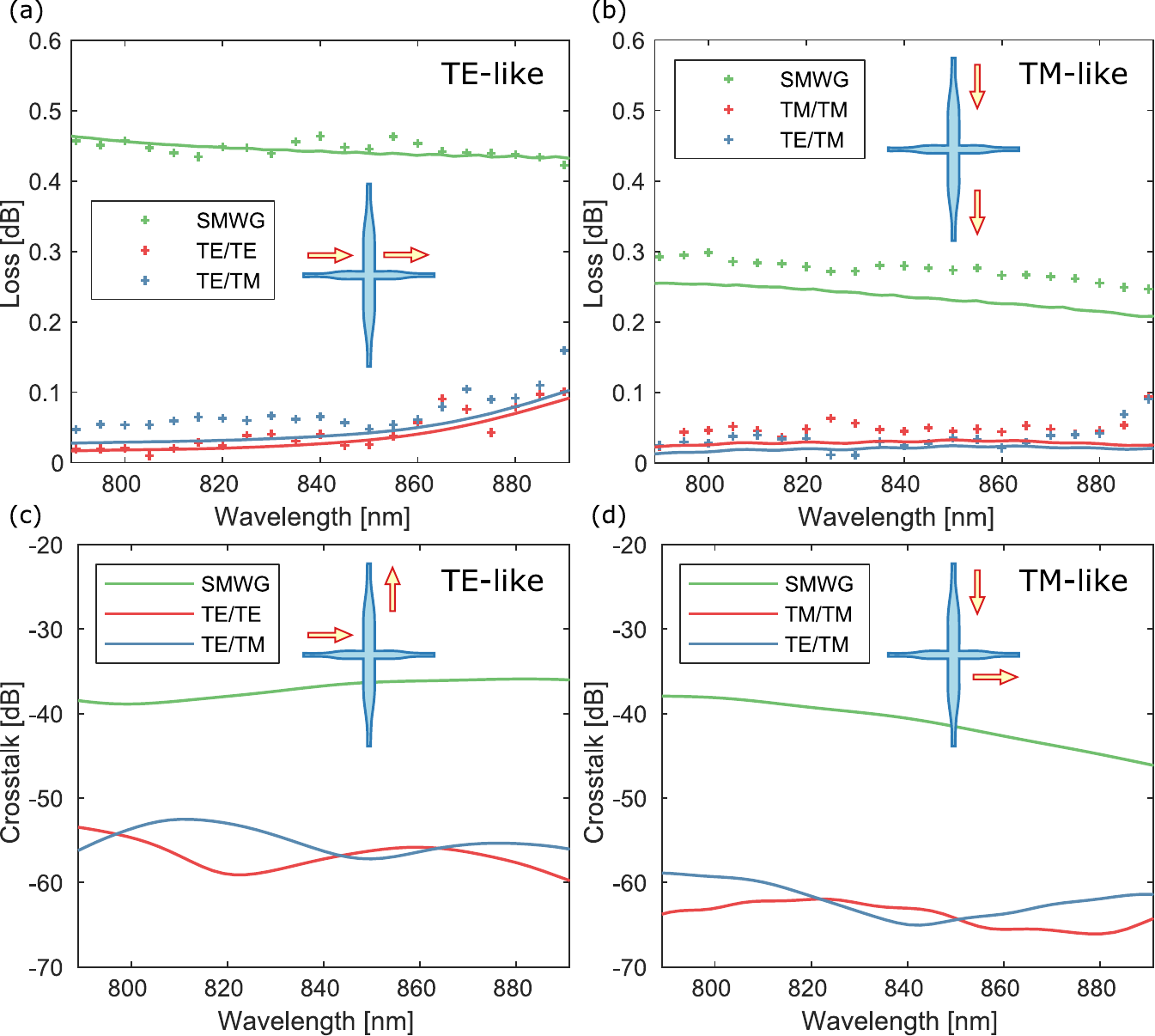}
    \caption{Comparison of loss and crosstalk of a single crossing between SMWG and optimized MMI crossings for light propagating in (a), (c)~TE-like polarization and (b), (d) TM-like polarization. The dots indicate the measurement results and the lines the 3D-FDTD simulations. For the SMWG crossing for TE-like polarization (a) a moving average of 100 measurement points was applied to smooth high frequency oscillations which can be attributed to interfering back reflection in the SMWG crossings. The insets depict the design of the TE/TM with arrows corresponding to the measurement setup.}
    \label{fig:loss_crosstalk_crossing}
\end{figure}

The different crossing configurations were then fabricated by ams AG on their SiN photonic platform \cite{Sagmeister.2018}. The measurements of the fabricated crossings were performed with cleaved polarization maintaining fibers coupled to the device under test via end facet coupling. The light was provided by a tunable Ti-sapphire laser. The light exiting the device was measured and referenced to the input power to normalize fluctuations of the laser. Due to the low losses of a single crossing multiple crossings were cascaded with an additional single mode waveguide sections in between. Figure~\ref{fig:microscopy_image} shows a false color infrared image of a part of the cascaded crossings.
The loss of a single crossing was determined by cascading both 40 and 80 crossings of each type. The measured transmission values were subtracted to remove the influence of the coupling loss, and divided by the difference in the number of crossings, i.e. by 40. The loss measurement results compared to the 3D-FDTD simulations can be seen in Fig.~\ref{fig:loss_crosstalk_crossing}(a) and Fig.~\ref{fig:loss_crosstalk_crossing}(b) for TE-like and TM-like polarization, respectively. The crossing losses from the simulations were calculated by normalizing the transmission from a power monitor placed after the crossing with a power monitor right in front of the crossing. The measurements show that losses for all optimized MMI crossing types over the observed bandwidths stay below \SI{0.16}{dB} and are lower compared to SMWG crossings by up to \SI{0.4}{dB} and \SI{0.25}{dB} for TE-like and TM-like polarization, respectively.

The crosstalk characteristics were investigated by 3D-FDTD simulations. An experimental evaluation was not possible due to the low crosstalk level of \SI{-40}{dB} even for the SMWG crossings, which can be seen from the theoretical predictions in Fig.~\ref{fig:loss_crosstalk_crossing}(c) and Fig.~\ref{fig:loss_crosstalk_crossing}(d) for TE-like and TM-like polarization, respectively. With the improved MMI crossings the crosstalk was further reduced to \SI{-60}{dB} according to simulations. However, qualitatively the 3D-FDTD simulations show an improved suppression of the crosstalk by \SI{20}{dB} of the optimized MMI crossing. In all aspects a clear improvement can be seen for both single and dual polarization crossings.

\begin{figure}
    \centering
    \includegraphics[width=13.3cm]{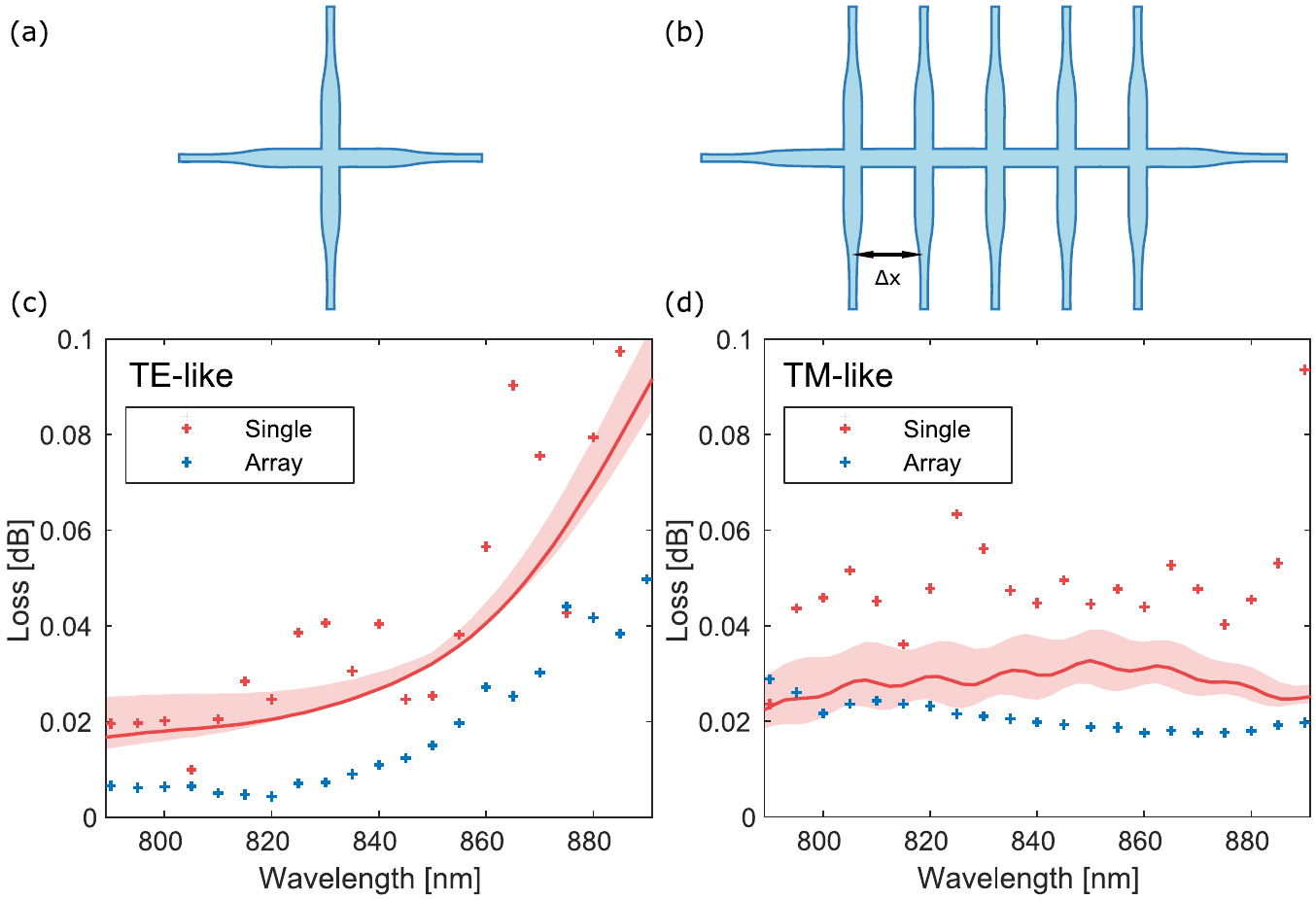}
    \caption{Schematic of a single crossing (a) compared to an array crossing (b) with multiple crossings in the same MMI structure with a constant distance. Corresponding loss measurement of a single crossing compared to arrayed crossings for (c)~the TE/TE configuration and (d)~the TM/TM configuration. The dots depict the measurement results and the continuous line the 3D-FDTD simulation of a single crossing. The shaded area indicates the loss distribution for a simulated waveguide width variation of up to $\pm \SI{20}{\nano \metre}$.
    }
    \label{fig:arrayed_crossing}
\end{figure}

In the case of multiple intersections on a single chip the crossing points can put in an array of crossings without the need of tapering down the MMI section to the single mode waveguide (see  Fig.~\ref{fig:arrayed_crossing}(b)), which results in the formation of Bloch waves along the MMI section \cite{Zhang.2013,Liu.2014}. This avoids losses accumulated during the tapering and allows for a smaller footprint. The optimum distance $\Delta \text{x}$ between arrayed crossings derived from 3D-FDTD simulations via a linear parameter sweep amounted to \SI{6.65}{\micro \metre} for the TE-like and \SI{13.05}{\micro \metre} for the TM-like polarization. For multiple crossings the array only has approximately \SI{40}{\%} of the length compared to cascaded crossings. To increase the accuracy for the loss measurement in array crossings the number of crossing points was increased to 100, 200 and 400. This made it possible to accurately determine losses below \SI{0.1}{dB} compared to a cascade of 80 crossings. Figure~\ref{fig:arrayed_crossing} compares the measured losses for a single crossing with those of an array crossing, showing an additional decrease in loss by \textasciitilde \SIrange{0.01}{0.02}{dB}. The loss in the arrayed crossings was determined via linear regression. The fact that the measured loss values of a single crossing tend to be above the simulated values, especially for TM-like polarization, can be attributed to the limitations of the measurement accuracy. With a difference of only 40 crossings for the single crossing measurement, the variation of the end facet coupling efficiency between individual test waveguides becomes more dominant. The simulated influence of the waveguide width on the loss indicates a relatively high tolerance with respect to width variations in fabrication.
\newpage
\section{Conclusion}
In conclusion we demonstrated low loss and low crosstalk silicon nitride waveguide crossings based on MMIs operating in the wavelength range of \SIrange{790}{890}{\nano \metre}. Across the wavelength range we observed losses below \SI{0.16}{dB}, which is a significant improvement compared to a standard single mode waveguide crossing of up to \SI{0.4}{dB} and \SI{0.25}{dB} for TE-like and TM-like polarization, respectively. Losses can be further reduced to \SI{0.05}{dB} and \SI{0.02}{dB} over the observed wavelength range for TE-like and TM-like polarization, respectively, using multiple crossings in an arrayed configuration without tapering down the MMI section. The crosstalk was shown to be around \SI{-60}{dB}, which is an improvement of about \SI{20}{dB} compared to standard single mode waveguide crossings. The crosstalk was evaluated via 3D-FDTD simulations due to the low crosstalk level and the resulting difficulty to differentiate it from light scattered from other positions on the PIC. With these results the presented crossing designs provide extended design freedom in on-chip routing for the corresponding silicon nitride waveguide PIC platform.
The optimized TE/TM crossing design further shows that dual polarization operation is possible on a single chip with high efficiency facilitating more versatile PIC layouts.
\section*{Funding}
European Union's Horizon 2020 research and innovation programme (688173).
\section*{Acknowledgments}
This research has received funding from the European Union's Horizon 2020 research and innovation program under grant agreement No 688173 (OCTCHIP).
\section*{Disclosure}
The authors declare no conflicts of interest.
%


\end{document}